\documentstyle[12pt]{article}

\begin{document}
\tolerance=5000
\def\pp{{\, \mid \hskip -1.5mm =}}
\def\cL{{\cal L}}
\def\be{\begin{equation}}
\def\ee{\end{equation}}
\def\bea{\begin{eqnarray}}
\def\eea{\end{eqnarray}}
\def\tr{{\rm tr}\, }
\def\nn{\nonumber \\}
\def\e{{\rm e}}
\def\D{{D \hskip -3mm /\,}}

\  \hfill
\begin{minipage}{3.5cm}
YITP-01-40 \\
NDA-FP-94 \\
May 2001 \\
\end{minipage}

\vfill

\begin{center}
{\large\bf  Holographic entropy and brane 
FRW-dynamics from AdS black hole 
in d5 higher derivative gravity}

\vfill

{\sc Shin'ichi NOJIRI}\footnote{nojiri@cc.nda.ac.jp},
{\sc Sergei D. ODINTSOV}$^{\spadesuit}$\footnote{
On leave from Tomsk State Pedagogical University, 
634041 Tomsk, RUSSIA. \\
odintsov@ifug5.ugto.mx, odintsov@mail.tomsknet.ru}, \\
and {\sc Sachiko OGUSHI}$^{\diamondsuit}$\footnote{
ogushi@yukawa.kyoto-u.ac.jp}\\

\vfill

{\sl Department of Applied Physics \\
National Defence Academy,
Hashirimizu Yokosuka 239-8686, JAPAN}

\vfill

{\sl $\spadesuit$
Instituto de Fisica de la Universidad de Guanajuato, \\
Lomas del Bosque 103, Apdo. Postal E-143, 
37150 Leon,Gto., MEXICO}

\vfill

{\sl $\diamondsuit$
Yukawa Institute for Theoretical Physics, \\
Kyoto University, Kyoto 606-8502, JAPAN}

\vfill

{\bf ABSTRACT}

\end{center}

Higher derivative bulk gravity (without Riemann tensor 
square term) admits 
AdS-Schwarzschild black hole as exact solution.  It is shown that 
induced brane geometry on such background is open, flat 
or closed FRW radiation dominated Universe.
Higher derivative terms contributions 
appear in the Hawking temperature, entropy  
and Hubble parameter via the redefinition of 
5-dimensional gravitational constant and AdS scale parameter. These
higher derivative terms do not destroy the AdS-dual 
description of radiation represented by strongly-coupled CFT. 
Cardy-Verlinde formula which expresses 
cosmological entropy  as square root from other parameters 
and entropies is derived in $R^2$ gravity. The corresponding 
cosmological entropy bounds are briefly discussed.

\newpage

\section{Introduction}

Brane-world physics (especially in the form of Randall-Sundrum 
proposal \cite{RS}) may provide new revolutionary ideas on
the structure of the early Universe and theory of fundamental 
interactions. Indeed, one can feel very incomfortable with 
the picture of observable Universe as tiny boundary of 
higher-dimensional (fundamental?) object
like black hole. Nevertheless, such situation is under 
current active discussion in high energy physics literature.
Moreover, it was realized recently \cite{SV} (for 
related works, see \cite{nooqr}) that brane equations of motion
are exactly Friedmann-Robertson-Walker(FRW) equations with 
radiation matter. This radiation matter plays the role of CFT
in AdS/CFT correspondence \cite{AdS}. This fact has direct 
interpretation as a consequence of holographic principle \cite{EV}.

Indeed, it was demonstrated in ref.\cite{EV} that FRW 
equation can be related with three different cosmological 
entropies bounds, giving kind of constraint among them. 
Furthermore, FRW equation can be rewritten in the form of
generalized Cardy-Verlinde formula \cite{Cardy,EV} relating 
cosmological entropy with the one of CFT filling the Universe. 
There was recently much activity on the studies of 
related questions \cite{others,nooqr}.

Our purpose in this work is further study of the CFT 
dominated Universe as the brane in the background of 
AdS Black Hole (BH). Our bulk theory is higher derivatives 
gravity which is known to possess AdS BH solution.
The interest in the higher derivatives bulk gravity 
is caused by the following. First of all, any effective 
stringy gravity includes higher derivative terms of 
different order. Second, from the point of view of 
AdS/CFT correspondence the $R^2$-terms give next-to-leading 
terms in large $N$ expansion \cite{Fa} as it was directly 
checked in the calculation of holographic conformal anomaly
from bulk $R^2$ gravity \cite{anom}. 
Third, higher derivative gravity
may serve as quite good candidate for the construction of
realistic brane-world cosmologies \cite{HDt,SNO}.

Our consideration is limited to specific model(c=0 model) 
of bulk $R^2$-gravity which does not contain the square of 
Riemann tensor in the action.
The attractive feature of this model is that it admits
Schwarzschild-Anti de Sitter black hole as exact solution 
in five dimensions.
Moreover, as it was demonstrated in ref.\cite{SNO} observable 
Universe could be the brane (boundary) of such BH. 
The extension to the case of c not equal zero (inclusion 
of Riemann tensor square) is evident\cite{SNO}, and 
qualitative conclusions are left the same.

The paper is organized as the following. In the next section 
the review of thermodynamics of AdS BH in d5 $R^2$ gravity is given.
Hawking temperature, horizon radius and entropy are presented in
the form suitable for later identification with the corresponding 
quantities in brane  FRW Universe. Section 3 is devoted to the discussion
of surface terms in higher derivative gravity, the surface counterterm on
AdS BH background is given. Brane dynamical equations are constructed.
In section 4 it is shown how induced geometry of brane takes the form
of open, flat or closed radiation dominated FRW equations.
 The role of higher derivative terms 
is analyzed.   The connection of higher derivative terms with AdS/CFT 
correspondence is mentioned. In the last section we construct 
Cardy-Verlinde formula in $R^2$ gravity. The cosmological entropy bounds are 
briefly discussed.

\section{AdS Black holes in bulk $R^2$-gravity and their  
thermodynamical properties}

Let us consider thermodynamics of AdS BH in bulk $R^2$-gravity.
The calculation of thermodynamical quantities like mass and entropy 
will be necessary to relate them with the corresponding ones in brane FRW 
Universe.
The general action of $d+1$-dimensional $R^2$ gravity is given by:
\be
\label{vi}
S=\int d^{d+1} x \sqrt{-\hat G}\left\{a \hat R^2 
+ b \hat R_{\mu\nu}\hat R^{\mu\nu}
+ c \hat R_{\mu\nu\xi\sigma}\hat R^{\mu\nu\xi\sigma}
+ {1 \over \kappa^2} \hat R - \Lambda \right\}\ .
\ee
The following conventions of curvatures are used
\bea
\label{curv}
R&=&g^{\mu\nu}R_{\mu\nu}\ , \nn
R_{\mu\nu}&=& -\Gamma^\lambda_{\mu\lambda,\nu}
+ \Gamma^\lambda_{\mu\nu,\lambda}
- \Gamma^\eta_{\mu\lambda}\Gamma^\lambda_{\nu\eta}
+ \Gamma^\eta_{\mu\nu}\Gamma^\lambda_{\lambda\eta}\ , \nn
\Gamma^\eta_{\mu\lambda}&=&{1 \over 2}g^{\eta\nu}\left(
g_{\mu\nu,\lambda} + g_{\lambda\nu,\mu} - g_{\mu\lambda,\nu} 
\right)\ .
\eea
When $a=b=c=0$, the action (\ref{vi}) becomes that of the 
Einstein gravity. 

When $c=0$ \footnote{For non-zero $c$ such S-AdS BH solution may be
constructed perturbatively\cite{SNO}. It is useful to establish the  
higher-derivative AdS/CFT correspondence\cite{NS1} and find the 
strong coupling limit of super Yang-Mills theory with two 
supersymmetries in next-to-leading order.}, 
Schwarzschild-anti de Sitter space is an exact solution:
\bea
\label{SAdS}
ds^2&=&\hat G_{\mu\nu}dx^\mu dx^\nu \nn
&=&-\e^{2\rho_0}dt^2 + \e^{-2\rho_0}dr^2 
+ r^2\sum_{i,j}^{d-1} g_{ij}dx^i dx^j\ ,\nn
\e^{2\rho_0}&=&{1 \over r^{d-2}}\left(-\mu + {kr^{d-2} \over d-2} 
+ {r^d \over l^2}\right)\ .
\eea
 The curvatures have the 
following form:
\be
\label{cv}
\hat R=-{d(d+1) \over l^2}\ ,\quad 
\hat R_{\mu\nu}= - {d \over l^2}\hat G_{\mu\nu}\ .
\ee
In (\ref{SAdS}), $\mu$ is the parameter corresponding to mass 
and the scale parameter $l$ is given by solving the following 
equation:
\be
\label{ll}
0={d^2(d+1)(d-3) a \over l^4} + {d^2(d-3) b \over l^4} \nn
- {d(d-1) \over \kappa^2 l^2}-\Lambda\ .
\ee
We also assume $g_{ij}$ expresses the Einstein manifold, 
defined by $r_{ij}=kg_{ij}$, where $r_{ij}$ is the Ricci tensor 
defined by $g_{ij}$ and $k$ is the constant. 
For example, if $k>0$ the boundary can be 4 dimensional 
de Sitter space (sphere when Wick-rotated), if $k<0$, anti-de Sitter 
space or hyperboloid, or if $k=0$, flat space. 
By properly normalizing the coordinates, one can choose $k=2$, $0$, 
or $-2$.

 The calculation of thermodynamical quantites like free 
energy $F$, the entropy ${\cal S}$ and the energy $E$ may be done 
with the help of the 
folllowing method  \cite{NS1}. After 
Wick-rotating the time variable by $t \to i\tau$, the
free energy $F$ can be obtained from the action $S$ (\ref{vi})
where the classical solution is substituted:
\bea
F=TS
\eea
Substituting eqs.(\ref{ll}) into (\ref{vi}) in the case of $d=4$ 
with $c=0$, one gets
\bea
\label{sr}
S &=& - \int d^{5}x \sqrt{-G} \left( {8 \over \kappa^2}
 - {320 a \over l^2} -{64 b \over l^2} \right) \nn
&=& -{V_{3} \over T}\int ^{\infty}_{r_{H}} dr r^{3}
\left( {8 \over \kappa^2}- {320 a \over l^2} -{64 b \over l^2} \right)
\eea
Here $V_{3}$ is the volume of 3d sphere and we assume $\tau$
has a period of ${1\over T}$.  The expression of $S$ contains
the divergence coming from large $r$. In order to subtract the
divergence, we regularize $S$ in (\ref{sr}) by cutting off the
integral at a large radius $r_{max}$ and subtracting the
solution with $\mu =0$ in a same way as in \cite{NS}:
\bea
S_{\rm reg}&=& 
-{V_{3} \over T}\left\{ \int ^{r_{max}}_{r_{H}} dr r^{3}
-e^{\rho(r=r_{max})-\rho(r=r_{max};\mu =0) }
\int ^{r_{max}}_{0} dr r^{3} \right\} \nn
&& \times 
\left( {8 \over \kappa^2 l^{2} }- {320 a \over l^{4}}
 -{64 b \over l^{4}} \right)
\eea
The factor $e^{\rho(r=r_{max})-\rho(r=r_{max};\mu =0)}$ is chosen
so that the proper length of the circle which corresponds to the
period ${1 \over T}$ in the Euclidean time at $r=r_{max}$
concides with each other in the two solutions.  Taking 
$r_{max} \to \infty$, one finds
\bea
F= V_{3}\left( {l^2 \mu \over 8} 
 - {r_H^4  \over 4}\right) \left( {8 \over \kappa^2}
 - {320 a \over l^2} -{64 b \over l^2} \right)
\eea
The horizon radius $r_{h}$ is given by solving 
the equation $e^{2\rho_0(r_H)}=0$ in (\ref{SAdS}):
\bea
\label{rh1}
r_{H}^{2}=-{k l^{2} \over 4} + {1\over 2}
\sqrt{{k^2 \over 4}l^{4}+ 4\mu l^{2} } \; .
\eea
The Hawking temperature $T_H$ is
\bea
\label{ht1}
T_H = {(e^{2\rho})'|_{r=r_{H}} \over 4\pi}
= {k \over 4\pi r_{H}} +{r_{H} \over \pi l^2} 
\eea
where $'$ denotes the derivative with respect to $r$. From 
the above equation (\ref{ht1}), $r_{H}$ can be rewritten  in terms of 
$T_{H}$ as
\bea
\label{rHTH}
r_{H}={1\over 2}\left( \pi l^{2} T_{H} \pm
\sqrt{(\pi l^{2} T_{H})^{2} -kl^{2}} \right)
\eea
In (\ref{rHTH}), the plus sign corresponds to $k=-2$ or $k=0$ 
case and the minus sign to $k=2$ case.\footnote{
When $k=2$, as we can see from (\ref{rh1}) and (\ref{ht1}), $r_H$, 
and also $T_H$, are finite in the limit of $l\rightarrow \infty$, 
which corresponds to the flat background. Therefore we need to 
choose the minus sign in (\ref{rHTH}) for $k=2$ case.}
 One can also rewrite $\mu $  using $r_{H}$ or $T_{H}$
from (\ref{rh1}) as follows:
\bea
\mu &=& {r_{H}^{4} \over l^{2}}+{kr_{H}^{2} \over 2}
=r_{H}^{2} \left( {r_{H}^{2} \over l^{2}}+{k \over 2} \right)\nn
&=& {1\over 4}\left( \pi l^{2} T_{H} \pm 
\sqrt{(\pi l^{2} T_{H})^{2} -kl^{2}} \right)^{2} \nn
&& \times \left({1\over 4 l^{2}}\left( \pi l^{2} T_{H} \pm
\sqrt{(\pi l^{2} T_{H})^{2} -kl^{2}} \right)^{2}
 +{k \over 2} \right)\ .
\eea 
Then we can rewrite $F$ using $T_{H}$ or $r_{H}$ as
\bea
F&=& -{V_{3} \over 32l^2}
\left( {8 \over \kappa^2}- {320 a \over l^2}
 -{64 b \over l^2} \right)
\left( \pi l^{2} T_{H} \pm 
\sqrt{(\pi l^{2} T_{H})^{2} -kl^{2}} \right)^{2} \\
&& \times \left(-\left( \pi l^{2} T_{H}+
\sqrt{(\pi l^{2} T_{H})^{2} -kl^{2}} \right)^{2} - kl^2 
\right) , \nn
&=& -{V_{3} \over 8}r_{H}^{2} \left( {r_{H}^{2} \over l^{2}}
 - {k \over 2} \right)\left( {8 \over \kappa^2} 
 - {320 a \over l^2} -{64 b \over l^2} \right) \; .
\eea
The entropy ${\cal S}$ and energy $E$ are 
\bea
\label{ent}
{\cal S }&=& -{dF \over dT_H}
=-{dF \over dr_{H}}{dr_{H} \over dT_H} \nn
&=& {V_{3} \over 16}\left({4 r_{H}^3 \over l^2} - kr_{H} \right)
\left(\pi l^{2} \pm { T_{H} \pi^{2} l^{4} \over \sqrt{\pi^{2}l^{4}
T_{H}^{2} -k l^{2} } } \right)
\left( {8 \over \kappa^2}- {320 a \over l^2} -{64 b \over l^2} \right)\nn
&=&{V_{3}\pi r_H^3 \over 2}
\left( {8 \over \kappa^2}- {320 a \over l^2}
 -{64 b \over l^2} \right)\nn
&=& {V_{3}\pi \over 16} 
\left( \pi l^{2} T_{H} \pm 
\sqrt{(\pi l^{2} T_{H})^{2} -kl^{2}} \right)^3 
\left( {8 \over \kappa^2}- {320 a \over l^2}
 -{64 b \over l^2} \right)\ .\\
\label{ener}
E&=&F+T{\cal S}\nn
&=& {3V_{3}\mu \over 8}
\left( {8 \over \kappa^2}- {320 a \over l^2}
 -{64 b \over l^2} \right)\ .
\eea
The above equations reproduce the standard Einstein theory results when 
$a=b=0$.  Note that  one can consider the limit of 
$l\rightarrow 0$, where the background spacetime becomes 
 flat Minkowski space. Since the scalar curvature and Ricci 
tensor vanishes in the flat Minkowski, we cannot derive the 
thermodynamical quantities by evaluating the action $S$, 
which vanishes, if we start with the flat Minkowski background 
from the begining. Then finite $l$ (or finite, non-vanishing 
cosmological constant) would give a kind of the regularization. 
Note also that above expressions will be used to get the cosmological 
entropy of brane FRW Universe.

\section{Surface terms in $R^2$-gravity}

Before considering the dynamics of the brane, we review the 
problem of the variational principle in the Einstein gravity, 
whose action is given by
\be
\label{E1}
S_{\rm E}={1 \over \kappa^2} \int d^{d+1} x \sqrt{-\hat G}
\hat R \ .
\ee
The scalar curvature contains the second order derivative 
of the metric tensor $\hat G_{\mu\nu}$ with respect the 
coordinates. Therefore if there is a boundary, which we denote 
by $B$, under the 
variation $\delta\hat G_{\mu\nu}$, $\delta S_{\rm E}$ 
contains, on the boundary, the derivative of 
$\delta\hat G_{\mu\nu}$ with respect to the coordinate 
perpendicular to the boundary, which makes the variational 
principle ill-defined. Therefore we need to add a surface 
term to the action, which is called  the Gibbons-Hawking 
surface term \cite{GB}. 
 Note that by using the definition of the curvature in 
(\ref{curv}), the Einstein action can be rewritten as 
\bea
\label{E2}
S_{\rm E}&=&{1 \over \kappa^2} \int d^{d+1} x \left[\sqrt{-\hat G}
\left(-\hat\Gamma_{\mu\lambda}^\eta \hat\Gamma_{\nu\eta}^\lambda 
+\hat\Gamma_{\mu\nu}^\eta\hat\Gamma_{\lambda\eta}^\lambda\right)
\hat G^{\mu\nu} \right.\nn
&& \left. + \hat\Gamma_{\mu\lambda}^\lambda
\partial_\nu\left(\sqrt{-\hat G}\hat G^{\mu\nu}\right)
 -  \hat\Gamma_{\mu\nu}^\lambda\partial_\lambda
 \left(\sqrt{-\hat G}\hat G^{\mu\nu}\right)\right] \nn
&& + {1 \over \kappa^2} \int_B d^d x \sqrt{-\hat g}
\left[-n_\nu \hat\Gamma_{\mu\lambda}^\lambda\hat G^{\mu\nu}
+ n_\lambda\hat\Gamma_{\mu\nu}^\lambda\hat G^{\mu\nu}\right]\ .
\eea
Here $n_\mu$ is the unit vector perpendicular to the boundary 
and $\hat g_{mn}$ is the boundary metric induced 
from $\hat G_{\mu\nu}$. 
Since the bulk part of the action does not contain the 
second order derivative of $\hat G_{\mu\nu}$ with respect to the 
coordinates, the variational principle becomes well-defined if 
we add the following boundary term to the Einstein action:
\be
\label{E3}
\tilde S_b=-{1 \over \kappa^2} \int_B d^d x \sqrt{-\hat g}
\left[-n_\nu \hat\Gamma_{\mu\lambda}^\lambda\hat G^{\mu\nu}
+ n_\lambda\hat\Gamma_{\mu\nu}^\lambda\hat G^{\mu\nu}\right]\ .
\ee
The action (\ref{E3}), however, breaks the general covariance 
in general. We should note, however, that
\be
\label{E4}
\nabla_\mu n_\nu=\partial_\mu n_\nu
 - \hat\Gamma_{\mu\nu}^\lambda n_\lambda \ ,\quad 
\nabla_\mu n^\nu=\partial_\mu n^\nu
 + \hat\Gamma_{\mu\lambda}^\nu n^\lambda 
\ee
and therefore 
\be
\label{E5}
-n_\nu \hat\Gamma_{\mu\lambda}^\lambda\hat G^{\mu\nu}
+ n_\lambda\hat\Gamma_{\mu\nu}^\lambda\hat G^{\mu\nu}
=\partial_\mu n^\mu + \hat G^{\mu\nu}\partial_\mu n_\nu 
 -2 \nabla_\mu n^\mu\ .
\ee
Then one can replace the boundary action $\tilde S_b$ with 
the Gibbons-Hawking one:
\be
\label{E6} 
S_{\rm GH}
={2 \over \kappa^2} \int_B d^d x \sqrt{-\hat g}\nabla_\mu n^\mu\ ,
\ee
 at least for the following metric
\be
\label{E7}
ds^2 = \left(1 + {\cal O}(q^2)\right)dq^2 
+ \hat g_{mn}(q,x^m) dx^m dx^n\ .
\ee
Here $q$ is the radial coordinate and the brane exists at $q=0$.
The difference $\partial_\mu n^\mu 
+ \hat G^{\mu\nu}\partial_\mu n_\nu$ between $\tilde S_b$ 
and $S_{\rm GH}$, which appears in (\ref{E5}), vanishes since 
both of $n_\mu$ and $n^\mu$ become constant vectors. 

Motivated by the above argument,  one assumes 
the surface terms in the following form 
\cite{NS,SNO}:\footnote{In \cite{NS,SNO}, more general 
counterterm corresponding to $S_b^{(1)}$ has been 
considered. From the requirements of the finiteness of bulk AdS and 
well-defined variational principle of the action, the 
counterterm has finally the form (\ref{sf1}). }
\bea
\label{sf}
S_{b}&=& S_{b}^{(1)}+S_{b}^{(2)} , \\
\label{sf1}
S_{b}^{(1)}&=& {2\over \tilde{\kappa}^2} \int d^{d}x \sqrt{\hat{g}} 
\nabla_{\mu}n^{\mu}  \, , \\
S_{b}^{(2)}&=& -\eta \int d^{d}x \sqrt{\hat{g}}\ .
\eea
The parameter $\eta$ (brane tension) which is usually 
free parameter in brane-world cosmology is not free any more 
and can be determined by the condition that 
the leading divergence of bulk AdS should vanish when one substitutes
the classical solution (\ref{SAdS}) 
into the action (\ref{vi}) with $c=0$ and into (\ref{sf}) or 
(\ref{sf1}):
\bea
\label{clactions}
S&=&\int d^4x r_0^d {1 \over d} \left\{{d^2(d+1)^2 a \over l^4}
+ {d^2(d+1)b \over l^4} - {d(d+1) \over \kappa^2 l^2}
 -\Lambda \right\} + {\cal O}\left(r_0^{d-1}\right) \nn
S_b&=&\int d^4x r_0^d \left\{ + {2d \over l^2 \tilde\kappa^2} 
 - {\eta \over l}\right\} + {\cal O}\left(r_0^{d-1}\right) \ .
\eea
Then one gets
\be
\label{eta1}
{\eta \over l}={d(d+1)^2 a \over l^4}
+ {d(d+1)b \over l^4} - {d+1 \over \kappa^2 l^2}
 - {\Lambda \over d} + {2d \over l^2 \tilde\kappa^2} 
\ee
or  deleting $\Lambda$ by using (\ref{ll})
\be
\label{eta2}
{\eta \over l}={4d(d+1)a \over l^4} + {4db \over l^4}
- {2 \over \kappa^2 l^2} + {2d \over l^2\tilde\kappa^2}\ .
\ee

Motivated with (\ref{E7}), we choose the metric in the following 
form: 
\be
\label{S1}
ds^2=dq^2 -\e^{\zeta(q,\tau)}d\tau^2 + \e^{\xi(q,\tau)}
g_{ij}dx^i dx^j\ .
\ee
Here $g_{ij}$ is the metric of the Einstein manifold as in 
(\ref{SAdS}). Then the curvatures etc. can be expressed as 
follows:
\bea
\label{S2}
\hat R_{qq} &=& - {\zeta_{,qq} \over 2} - {\zeta_{,q}^2 \over 4}
+ (d-1)\left( -{\xi_{,qq} \over 2}
 - {\xi_{,q}^2 \over 4}\right)\ ,\nn
\hat R_{\tau\tau} &=& \e^\zeta\left({\zeta_{,qq} \over 2} 
+ {\zeta_{,q}^2 \over 4} + {(d-1) \zeta_{,q}\xi_{,q} \over 4}
\right) \nn
&& + (d-1)\left( -{\xi_{,\tau\tau} \over 2}
 - {\xi_{,\tau}^2 \over 4} + {\zeta_{,\tau}\xi_{,\tau} 
 \over 4}\right)\ ,\nn
\hat R_{q\tau}&=&R_{\tau q}= (d-1)\left\{ - {\xi_{,q\tau} \over 2} 
- {\xi_\tau \left(\xi_q - \zeta_q\right) \over 4}\right\}\ ,\nn
\hat R_{ij}&=& g_{ij}\left[ k + \e^{-\zeta + \xi} 
\left( {\xi_{,\tau\tau} \over 2}
 + {(d-1)\xi_{,\tau}^2 \over 4} - {\zeta_{,\tau}\xi_{,\tau} 
 \over 4}\right) \right. \nn
&& \left. + \e^\xi \left(-{\xi_{,qq} \over 2}
 - {(d-1)\xi_{,q}^2 \over 4} - {\zeta_{,q}\xi_{,q} 
\over 4} \right)\right]\ ,\\
&& \mbox{other components of Ricci tensor}=0 \ ,\nn
R&=& (d-1)k\e^{-\xi} + (d-1)\e^{-\zeta}\left(
\xi_{,\tau\tau} + d\xi_{,\tau}^2 - {\zeta_{,\tau}\xi_{,\tau} 
 \over 2}\right) \nn
&& + (d-1)\left(- \xi_{,qq} - {d\xi_{,q}^2 \over 4}
 - {\zeta_{,q}\xi_{,q} \over 2} \right)
 - \zeta_{qq} - {\zeta_{,q}^2 \over 2}\ ,\nn
\nabla_\mu n^\mu &=& {\zeta_{,q} \over 2} + {(d-1)\xi_{,q} 
\over 2}\ .
\eea
 Assume the bulk solution has the form of (\ref{cv}). 
Then the variation of the action at the boundary 
has the following form:
\bea
\label{S3}
&& \delta S + \delta S_b \nn
&& = \int d^{d}x \sqrt{\hat{g}}\left[
\left( {1 \over \tilde \kappa^2} - {1 \over \kappa^2} 
+ {2d(d+1)a \over l^2} + {2db \over l^2}\right)
\left( \delta\zeta_{,q} + (d-1) \delta\xi_{,q} \right) 
\right. \nn
&& - {1 \over 2}\left( - {1 \over \kappa^2} 
+ {2d(d+1)a \over l^2} + {2db \over l^2}\right)
\left( \zeta_{,q} + (d-1) \xi_{,q} \right) 
\left( \delta\zeta + (d-1) \delta\xi \right) \nn
&& \ + (d-1)\left\{\left(- {1 \over \kappa^2} 
+ {2d(d+1)a \over l^2} + {2db \over l^2}\right)\left( 
{\zeta_{,q} \over 2} + {d \xi_{,q} \over 2} \right) \right. \nn
&& \ \left.+ {1 \over 2\tilde\kappa^2}\left(\zeta_{,q} + (d-1)\xi_{,q}
\right) - {\eta \over 2}\right\} \delta\xi \nn
&& \ + \left\{\left( - {1 \over \kappa^2} 
+ {2d(d+1)a \over l^2} + {2db \over l^2}\right)\left( 
\zeta_{,q}  + {(d-1) \xi_{,q} \over 2} \right) \right. \nn
&& \left.\left. + {1 \over 2\tilde\kappa^2}\left(\zeta_{,q} + (d-1)\xi_{,q}
\right) - {\eta \over 2}\right\} \delta\zeta \right]\ .
\eea
Then taking $\tilde\kappa$ as in the previous 
work \cite{SNO}, 
\be
\label{S4}
{1 \over \tilde\kappa^2}= {1 \over \kappa^2} 
 - {2d(d+1) a \over l^2 } - {2d b \over l^2}\ ,
\ee
the coefficients in front of $\delta\zeta_{,q}$ and 
$\delta\xi_{,q}$ vanish. Then the variation of the 
action becomes simple:
\bea
\label{S5}
\delta S + \delta S_b 
&=& \int d^{d}x \sqrt{\hat{g}}\left[(d-1)\left(  
{1 \over 2\tilde\kappa^2}\left(\zeta_{,q} + (d-2)\xi_{,q}
\right) - {\eta \over 2}\right)
\delta\xi  \right. \nn
&& \left. + \left( 
{d-1 \over 2\tilde\kappa^2}\xi_{,q} - {\eta \over 2}\right)
\delta\zeta \right]\ .
\eea
The dynamical brane equations look as
\be
\label{S6}
\left.\left(\zeta_{,q} + (d-1)\xi_{,q}\right)\right|_{q=0}
= \left.(d-1)\xi_{,q}\right|_{q=0}
= \tilde\kappa^2 \eta\ .
\ee
In Eqs.(\ref{S5}) and (\ref{S6}), it is supposed $\tilde\kappa$ 
is given by (\ref{S4}). 
Combining (\ref{eta2}) and (\ref{S4}), we find 
\be
\label{B14}
\eta = {2 (d-1)\over l\tilde\kappa^2}\ .
\ee
Then (\ref{S6}) can be rewritten as
\be
\label{S6b}
\left.\left(\zeta_{,q} + (d-1)\xi_{,q}\right)\right|_{q=0}
= \left.(d-1)\xi_{,q}\right|_{q=0}
={2(d-1) \over l}\ .
\ee
Especially when $\e^\xi=\e^\zeta=l^2\e^{2A}$, where the metric 
is given by
\be
\label{metric}
ds^2=dq^2 + l^2\e^{2A}\left(-d\tau^2 + \sum_{i,j=1}^{d-1} 
g_{ij}dx^i dx^j\right)\ ,
\ee
 one obtains 
\be
\label{A}
\left.A_{,q}\right|_{q=0}
= {1 \over l}\ .
\ee
The contribution of the higher derivative terms come 
through $l$ (\ref{ll}). We should note, however, that 
${1 \over l}$ is finite even if bulk cosmological constant 
$\Lambda$ vanishes.

When $d=4$, Eq.(\ref{S4}) has the following form:
\be
\label{B16}
{1 \over \tilde\kappa^2}={1 \over \kappa^2} - {40a \over l^2} 
 - {8b \over l^2}\ .
\ee
It is non-zero even if bulk Einstein action is absent.
The entropy (\ref{ent}) and the energy (\ref{ener}) have 
the following form:
\bea
\label{ent2}
{\cal S }&=& {4V_{3}\pi r_H^3 \over \tilde\kappa^2}\ ,\\
\label{ener2}
E&=&{3V_{3}\mu \over \tilde\kappa^2}\ .
\eea
Therefore the corrections from the higher derivative terms 
appear through the redefinition of  gravitational coupling $\kappa$ to 
$\tilde\kappa$ through (\ref{S4}) or (\ref{B16}) and the 
length scale $l$ given by (\ref{ll}). As one sees in a moment 
the above entropy is cosmological entropy of FRW Universe. 

\section{The FRW equation of the brane cosmology  from 
$R^2$-gravity}

 Let us rewrite the metric (\ref{SAdS}) of Schwarzschild-anti 
de Sitter space in a form of (\ref{S1}) or (\ref{metric}). 
If one chooses coordinates $(q,\tau)$ as
\bea
\label{cc1}
&& l^2\e^{2A-2\rho_0}A_{,q}^2 - \e^{2\rho_0} t_{,q}^2 = 1 \ ,
\quad l^2\e^{2A-2\rho_0}A_{,q}A_{,\tau}
 - \e^{2\rho_0}t_{,q} t_{,\tau}= 0 \nn
&& l^2\e^{2A-2\rho_0} A_{,\tau}^2 - \e^{2\rho_0} t_{,\tau}^2 
= -l^2\e^{2A}\ .
\eea
the metric takes the form (\ref{metric}). Here $r=l\e^A$.
Furthermore choosing a coordinate $\tilde t$ by 
$d\tilde t = l\e^A d\tau$, 
the metric on the brane takes FRW form: 
\be
\label{e3}
ds_{\rm brane}^2= -d \tilde t^2  + l^2\e^{2A}
\sum_{i,j=1}^{d-1}g_{ij}dx^i dx^j\ .
\ee
By solving Eqs.(\ref{cc1}), we have
\be
\label{e4}
H^2 = A_{,q}^2 - {\e^{2\rho_0}\e^{-2A} \over l^2}\ .
\ee
Here the Hubble constant $H$ is defined by
$H={dA \over d\tilde t}$. 
Then using (\ref{SAdS}) and (\ref{A}),  one obtains the 
following equation:
\be
\label{e5}
H^2={1 \over l^2} - {1 \over r^d}
\left(-\mu + {kr^{d-2} \over d-2}  + {r^d \over l^2}\right)
= - {k \over (d-2)r^2} + {\mu \over r^d}
\ .
\ee
Especially, when $k=d-2>0$, the spacial part of 
the brane has the shape of the $(d-1)$-dimensional sphere 
and $r$ can be regarded 
as the radius of the spacial part of the brane universe. 

Eq.(\ref{e5}) can be rewritten in the form of the 
FRW equation (compare with \cite{SV}):
\bea
\label{F1}
H^2 &=& - {k \over (d-2)r^2} + {\kappa_d^2 \over (d-1)(d-2)}
{\tilde E \over V}\ ,\nn
{\tilde E}&=&{(d-1)(d-2) \mu V_{d-1} \over \kappa_d^2 r}\ ,\nn
V&=&r^{d-1}V_{d-1}\ .
\eea
Here $V_{d-1}$ is the volume of the $(d-1)$-dimensional sphere 
with a unit radius  and $\kappa_d$ is the $d$-dimensional 
gravitational coupling, which is given by
\be
\label{F2}
\kappa_d^2={2 \tilde\kappa^2 \over l}\ .
\ee
By differentiating Eq.(\ref{F1}) with respect to $\tilde t$, 
since $H={1 \over r}{dr \over d\tilde t}$, we obtain the 
second FRW equation
\bea
\label{2FR1}
\dot H &=& - {\kappa_d^2 \over 2(d-2)} \left({\tilde E \over V} 
+ p\right) + {k \over (d-2)r^2}\ ,\nn
p&=&{(d-2)\mu \over r^d \kappa_d^2}\ .
\eea
Here $p$ can be regarded as the pressure of the matter on 
the brane.

In case of the Einstein gravity ($a=b=c=0$ in (\ref{vi})), 
the equation corresponding to (\ref{F2}) has the form 
\be
\label{F2E}
\kappa_{{\rm (Ein)}d}^2={2 \kappa^2 \over l}\ .
\ee
Then the effects of the higher derivative terms appear 
through the redefinition of $\kappa^2$ to $\tilde\kappa^2$. 
 Note, however, that one can obtain (\ref{F2}) even  
without the Einstein term (${1 \over \kappa^2}=0$).

When $d=4$, by using (\ref{ener2}) we find
\be
\label{F3}
\tilde E={l \over r}E\ .
\ee
Note that when $r$ is large, the metric 
(\ref{SAdS}) has the following form:
\be
\label{eq13} 
ds_{\rm AdS-S}^2 \rightarrow {r^2 \over l^2}\left(-dt^2 
+ l^2 \sum_{i,j}^{d-1} 
g_{ij}dx^i dx^j)\right)\ ,
\ee
which tells that the CFT time $\tilde t$ is equal 
to the AdS time $t$ times the factor ${r \over l}$:
\be
\label{eq14}
t_{\rm CFT}={a \over l}t\ .
\ee
Therefore Eq.(\ref{F3}) expresses that 
the energy in CFT is related 
with the energy $E$ in AdS by a factor ${l \over r}$ \cite{SV}. 

Eq.(\ref{F1}) or (\ref{F3}) tells that $\tilde E$ 
scales as $\tilde E \rightarrow \lambda^{-1}\tilde E$ when one 
scales the radius of the brane universe as 
$r\rightarrow \lambda r$. From Eqs.(\ref{F1}) and (\ref{2FR1}), 
we find 
\be
\label{trace1}
0=-{\tilde E \over V} + (d-1)p\ ,
\ee
which tells that the trace of the energy-stress tensor coming 
from the matter on the brane vanishes:
\be
\label{trace2}
{T^{{\rm matter}\ \mu}}_\mu=0\ .
\ee
Therefore the matter on the brane can be 
regarded as the radiation, i.e., the massless fields. 
 In other words, field theory on the brane should be conformal one as in case 
of Einstein brane \cite{SV}. This supports the claim that even for $c=0$ 
case, the higher derivative terms in bulk gravity correspond to 
next-to-leading corrections in AdS/CFT set-up.
 
 Using (\ref{S4}) and (\ref{F2}), we can rewite $\tilde E$  as
\be
\label{tE}
{\tilde E}={2(d-1)(d-2) \kappa^2\mu V_{d-1} \over r
\left( 1 - {2d(d+1)\kappa^2 a \over l^2 }
 - {2d b\kappa^2 \over l^2}\right)}\ .
\ee
 Assuming AdS/CFT correspondence, the higher derivative 
terms in (\ref{vi}) correspond to the $1/N$ corrections 
in the large $N$ limit of some gauge theory, which could 
be a CFT on the brane. For $\kappa^2 a$, 
$\kappa^2 b\ll 1$,  one can rewrite (\ref{tE}) as
\be
\label{tE2}
{\tilde E}\sim {2(d-1)(d-2) \kappa^2\mu V_{d-1} \over r}
\left( 1 + {2d(d+1)\kappa^2 a \over l^2 }
+ {2d b\kappa^2 \over l^2}\right)\ .
\ee
Then the parameters $a$ and $b$ in 
(\ref{tE2}) could express the $1/N$ correction of the 
next-to-leading order of $1/N$ expansion.

We now restrict to $d=4$ case and let the entropy ${\cal S}$ 
of CFT on the brane is given 
by the entropy (\ref{ent}) of the AdS$_5$ black hole. 
If the total entropy ${\cal S}$ is constant during the 
cosmological evolution, the entropy density $s$ is given 
by (see \cite{SV})
\be
\label{e20}
s={{\cal S} \over r^3 V_3}
= {4V_{3}\pi r_H^3 \over \tilde\kappa^2}
{8\pi r_H^3 \over l\kappa_4 r^3}\ .
\ee
Here  Eq.(\ref{ent}) is used. 
The temperature $T$ on the brane 
is different from Hawking temperature $T_H$ by the factor 
${l \over r}$:
\be
\label{e22}
T={l \over r}T_H
={r_H \over \pi r l} + {kl \over 4\pi r r_H}\ .
\ee
Then especially when $r=r_H$
\be
\label{e23}
T={1 \over \pi l} + {k \over 4\pi r_H^2}\ .
\ee
If the energy and entropy are purely extensive, the 
quantity $\tilde E + pV - TS$ vanishes. But in general, 
this quantity does not vanish and  one can define the 
Casimir energy $E_C$ by
\be
\label{EC1}
E_C=3\left(\tilde E + pV - TS\right)\ .
\ee
(The factor $3$ is replaced by $d-1$ in the general dimensions). 
Then by using, Eqs.(\ref{F1}), (\ref{2FR1}) for $d=4$ and 
Eqs.(\ref{e20}), (\ref{e22}), we find  
\bea
\label{EC2}
E_C&=&{6k r_H^2 V \over \kappa_4^2 r^4} \nn
&=& {6k r_H^2 V_3 \over \kappa_4^2 r} \nn
&=& {3l k r_H^2 V_3 \over \kappa_4^2 r}\left(1
 - {40a \kappa^2 \over l^2}
  - {8b\kappa^2 \over l^2} \right) \ .
\eea
When $k=0$, the Casimir energy vanishes. 
When $a$ and $b$ are small, $E_C$ is positive (negative) when 
 $k=2$ ($k=-2$) but if $a$ or $b$ is large and positive, 
$E_C$ can be negative (positive) even if  $k=2$ ($k=-2$).
In case of absence of higher derivative terms corrections the above
Casimir energy coincides with the one calculated in ref.\cite{youm}. 

Of course, as horizon radius has higher derivative terms corrections, 
 the quantities found in this section are different
 from the ones in Einstein theory.
This finishes our discussion of open, flat or closed
 radiation-dominated FRW Universe equations
as it follows from induced geometry of brane living in d5 AdS BH (solution 
of bulk $R^2$ gravity).

\section{Cardy-Verlinde formula in $R^2$-gravity}

In \cite{EV}, it was shown that the FRW equation in 
$d$-dimensions can be regarded as a $d$-dimensional analogue of
the Cardy formula of 2d conformal field theory (CFT) \cite{Cardy}:
\be
\label{CV1}
\tilde {\cal S}=2\pi \sqrt{
{c \over 6}\left(L_0 - {k \over d-2}{c \over 24}\right)}\ .
\ee
In the present case, identifying
\bea
\label{CV2}
{2\pi \tilde E r \over d-1} &\Rightarrow& 2\pi L_0 \ ,\nn
{(d-2)V \over \kappa_d^2 r} &\Rightarrow& {c \over 24} \ ,\nn
{4\pi (d-2)HV \over \kappa_d^2} &\Rightarrow& \tilde {\cal S}\ ,
\eea
the FRW-like equation (\ref{F1}) has the form (\ref{CV1}). 

The total entropy of the universe could be conserved in the 
expansion. Then one can evaluate holographic (Hubble) entropy 
$\tilde{\cal S}$ in (\ref{CV2}) 
when the brane crosses the horizon $r=r_H$. When $r=r_H$, 
Eq.(\ref{F1}) tells that 
\be
\label{CV3}
H=\pm {1 \over l}\ .
\ee
Here the plus sign corresponds to the expanding brane universe 
and the minus one to the contracting universe.  Taking 
the expanding case and using (\ref{CV2}), we find
\be
\label{CV4}
\tilde{\cal S}={4\pi (d-2) V \over l\kappa_d^2}
={2\pi (d-2) r_H^{d-1} V_{d-1} \over \tilde\kappa^2}\ .
\ee
Especially for $d=4$, the entropy $\tilde{\cal S}$ is 
identical with ${\cal S}$ in (\ref{ent}), which is nothing 
but the black hole entropy
\be
\label{CV5}
\tilde{\cal S}={\cal S}\ .
\ee
 Generally, one gets
\bea
\label{CV6}
&\tilde{\cal S}>{\cal S} \quad &\mbox{if}\quad Hl>1 \nn
&\tilde{\cal S}<{\cal S} \quad &\mbox{if}\quad Hl<1 \ .
\eea

As in \cite{EV}, if defining the Bekenstein entropy $S_B$ 
and the Bekenstein-Hawking entropy $S_{BH}$ by
\be
\label{CV7}
S_B={2\pi \over d-1}Er=2\pi L_0\ ,
\quad S_{BH}={4\pi(d-2)V \over \kappa_d^2 r}={\pi \over 6}c\ ,
\ee
we have
\be
\label{CV8}
\tilde{\cal S}^2 = 2S_B S_{BH} - {k \over d-2}S_{BH}^2 \ ,
\ee
 It is interesting that this equation becomes $k$-dependent as in \cite{youm}.
$R^2$ gravity corrections are hidden in the entropies.
 One can also define the Casimir entropy $S_C$ by
\be
\label{CV9}
S_C={4\pi r \over (d-1)k} E_C\ .
\ee
When $d=4$,  using (\ref{EC2}) one gets
\be
\label{CV10}
S_C={8\pi r_H^2 V \over \kappa_4^2 r^3} \ .
\ee
By comparing the above expression $S_{BH}$ in (\ref{CV7}), 
we find 
\be
\label{CV11}
S_C>S_{BH}\ \mbox{if}\ r<r_H\ \mbox{and}\ 
S_C<S_{BH}\ \mbox{if}\ r>r_H\ .
\ee

We now stress again that compared with the Einstein gravity case in 
\cite{SV}, the corrections from the higher derivative terms 
always appear through the redefinition of gravitational coupling 
$\kappa$ to 
$\tilde\kappa$ via (\ref{S4}) or (\ref{B16}) and when the 
length scale $l$ is given by (\ref{ll}). From the viewpoint of 
AdS/CFT correspondence, the higher derivative terms in (\ref{vi}) 
correspond to the $1/N$ corrections in the large $N$ limit of 
some gauge theory, which could be a CFT on the brane.  
Then Eqs.(\ref{F3}) and (\ref{CV5}) would tell that AdS/CFT 
correspondence could be valid in the next-to-leading order of 
the $1/N$. 

Of course, it is interesting to investigate the $c\neq 0$ case.
It may include also the situation when bulk gravity is Gauss-Bonnet one.
When $c\neq 0$ case, however, the Schwarzschild-AdS space 
in (\ref{SAdS}) is not an exact solution. Then  one should treat 
$c$ as a perturbation. Although the calculation becomes tedious, 
it is straightforward and does not change the qualitative conclusions. 
It will be investigated elsewhere.

In summary, the emergence of brane FRW Universe dynamics as well as 
appearence of  
holographic cosmological entropy from AdS BH in d5 $R^2$ gravity 
is demonstrated. Despite the presence of parameters from 
higher derivative bulk terms,
the radiation is represented by a strongly coupled CFT as 
it happens in purely Einstein theory. 
Cardy-Verlinde formula for cosmological entropy in $R^2$ gravity is derived. 
The corresponding cosmological entropy bounds are briefly discussed.
Our study indicates to affirmative answer to the 
question of \cite{SNO}:
Can we live on the brane in SAdS black hole?

\section*{Acknoweledgements} 

The work by SDO was supported in part by 
CONACyT (CP,Ref.990356 and grant 28454E).
The work of S.O. was supported in part by the Japan Society for the
Promotion of Science under the Postdoctoral Research Program No. 11-05921.


\begin{thebibliography}{99}
\bibitem{RS} L. Randall and R. Sundrum,
 {\sl Phys.Rev.Lett.} {\bf 83} (1999) 3370, hep-th/9905221;
 {\sl Phys.Rev.Lett.}  {\bf 83} (1999) 4690, hep-th/9906064.
\bibitem{SV} I. Savonije and E. Verlinde, hep-th/0102042.
\bibitem{nooqr} S. Nojiri and S.D. Odintsov, hep-th/0102042;
Y.S. Myung, hep-th/0103241;
N.J. Kim, H.W. Lee and Y.S. Myung, hep-th/0104159;
\bibitem{AdS}  J.M. Maldacena, 
{\sl Adv.Theor.Math.Phys.} {\bf 2} (1998) 231;
E. Witten, {\sl Adv.Theor.Math.Phys.} {\bf 2} (1998) 253;
S. Gubser, I. Klebanov and A. Polyakov, {\sl Phys.Lett.} 
{\bf B428} (1998) 105;
O. Aharony, S. Gubser, J. Maldacena, H. Ooguri and Y. Oz,
{\sl Phys.Repts.} {\bf 323} (2000) 183.
\bibitem{EV} E. Verlinde, hep-th/0008140.
\bibitem{Cardy} J.L. Cardy, {\sl Nucl.Phys.} {\bf B270} 
(1986) 967. 
\bibitem{others}
D. Kutasov and F. Larsen, hep-th/0009244;
F.-L. Lin, hep-th/0010127;
S. Nojiri and S.D. Odintsov, hep-th/0011115, IJMPA, to appear;
B. Wang, E. Abdalla and R.-K.Su, hep-th/0101073
D. Klemm, A. Petkou and G. Siopsis, hep-th/0101076;
Y.S. Myung, hep-th/0102184;
R. Brustein, S. Foffa and G. Veneziano, hep-th/0101083;
R.-G. Cai, hep-th/0102113;
A. Biswas and S. Mukherji, hep-th/0102138;
D. Birmingham and S. Mokhtari, hep-th/0103108
D. Klemm, A. Petkou, G. Siopsis and D. Zanon, hep-th/0104141; 
D. Youm, hep-th/0105036;
S. Nojiri, O. Obregon, S.D. Odintsov, H. Quevedo and M.P. Ryan, hep-th/0105052;
 R.-G. Cai, Y.S. Myung and N. Ohta, hep-th/0105070.
\bibitem{Fa} A. Fayyazuddin and M. Spalinski,hep-th/9805096,
 {\it Nucl.Phys.}{\bf B535} (1998) 219;
O. Aharony, A. Fayyazuddin and J. Maldacena, hep-th/9806159,
 {\it JHEP}{\bf 9807} (1998) 013;
\bibitem{anom} S. Nojiri and S.D. Odintsov, hep-th/9903033,
 {\it Int.J.Mod.Phys.Lett.}{\bf A15} (2000) 413; hep-th/9910113,
 {\it Mod.Phys.Lett.} {\bf A15} (2000) 1043;
M. Blau, K.S. Narain and E. Gava, hep-th/9904179,
 {\it JHEP} {\bf 9909} (1999) 018;
A. Bilal and C.-S. Chu, hep-th/9907106, {\it Nucl.Phys.} {\bf B562} (1999) 181;
M. Fukuma, S. Matsuura and T. Sakai, hep-th/0103187.
\bibitem{HDt}
J.E. Kim, B. Kyae and H.M. Lee, hep-ph/9912344,{\it Phys.Rev.} {\bf D62}
(2000)045013; J.E. Kim and H.M. Lee, hep-th/0010093;
N. Deruelle and T. Dolezel, gr-qc/0004021, {\it Phys.Rev.}
{\bf D62} (2000)103502;
S. Nojiri and S.D. Odintsov, hep-th/0006232, {\it JHEP} {\bf 0007} (2000) 049;
N.E. Mavromatos and J. Rizos, hep-th/0008074, {\it Phys.Rev.}{\bf D62} (2000)
124004; I. Neupane, hep-th/0008190, {\it JHEP} {\bf 0009} (2000) 040;
K. Kashima, hep-th/0010286;
M. Giovannini, hep-th/0011153;
S. Mukohyama, hep-th/0101038, {\it Phys.Rev.}{\bf D63} (2001) 104025;
Y.M. Cho, I. Neupane and P.S. Wesson, hep-th/0104227.
\bibitem{SNO} S. Nojiri, S.D. Odintsov, hep-th/0007205.
{\it Phys.Lett. }{\bf B493}(2000) 153;
S. Nojiri, S.D. Odintsov and S. Ogushi, hep-th/0010004, PTP, to appear.
\bibitem{NS1} S. Nojiri, S.D. Odintsov, hep-th/9908065,
{\it Phys.Lett. }{\bf B471} (1999) 155.
\bibitem{NS} S. Nojiri, S.D. Odintsov, hep-th/9911152,
{\it Phys.Rev.}{\bf D 62}(2000) 064018.
\bibitem{GB} G. Gibbons and S.W. Hawking, 
{\it Phys.Rev.}{\bf D15} (1977)2752. 
\bibitem{youm} D. Youm, hep-th/0105093. 
\end{thebibliography}
\end{document}